\documentstyle[preprint,aps,floats,epsfig]{revtex}

\tightenlines

\def\appendix{\par
 \setcounter{section}{0}
 \setcounter{subsection}{0}
 \def\thesection{Appendix \Alph{section}}
 \def\thesubsection{\Alph{section}.\arabic{subsection}}
 \def\theequation{\Alph{section}.\arabic{equation}}
 \setcounter{equation}{0}}
\newcommand{\be}{\begin{equation}}
\newcommand{\ee}{\end{equation}}
\newcommand{\bear}{\begin{eqnarray}}
\newcommand{\eear}{\end{eqnarray}}

\begin{document}
\preprint{USM-TH-108}
\draft
\title{Resummations with renormalon effects
for the hadronic vacuum polarization contribution to the muon (g-2)}

\author{Gorazd Cveti\v c$^1$\footnote{cvetic@fis.utfsm.cl},
Taekoon Lee$^2$\footnote{tlee@muon.kaist.ac.kr},
and Iv\'an Schmidt$^1$\footnote{ischmidt@fis.utfsm.cl}}
\address{$^1$Department of Physics,
Universidad T\'ecnica Federico Santa Mar\'{\i}a,
Valpara\'{\i}so, Chile\\
$^2$Department of Physics, KAIST, Daejon 305-701, Korea}
\date{\today}

\maketitle

\begin{abstract}
The hadronic vacuum polarization contribution to the muon $(g-2)$ value is
calculated by considering a known dispersion integral
which involves the $R_{e^+ e^-}(s)$ ratio.
The theoretical part stemming from the region below
1.8 GeV is the largest contribution in our approach, and
is calculated by using a contour integral involving
the associated Adler function $D(Q^2)$. In the resummations, we 
explicitly take into account the exactly known renormalon singularity 
of the leading infrared renormalon in the usual and in the modified Borel 
transform of $D(Q^2)$, and map further away from the origin the other
renormalon singularities by employing judiciously chosen
conformal transformations.
The renormalon effect increases the predicted value of the  
hadronic vacuum polarization contribution to the muon $(g-2)$, and 
therefore diminishes the difference 
between the recently measured and the SM/QCD-predicted value of $(g-2)$.
It is also shown that the total QED correction to the hadronic 
vacuum polarization is very small, about  0.06~\%. 
\end{abstract}

\section{Introduction}

The new precise measurement of the muon anomalous magnetic moment
$a_{\mu} \equiv (g-2)/2$ \cite{Brown:2001mg} allows for detailed
testing of the standard model, and therefore for the possibility
of looking into physics beyond the standard model as well. In
fact, comparison of the experimental result with some theoretical
calculations shows a $ 2.6 \sigma$ difference \cite{Brown:2001mg}.
This has been suggested as the appearance of new physics.\footnote{
For an (incomplete) list of works investigating the possibility
to explain and/or to find implications of this difference
within various frameworks of new physics, see: Refs.~\cite{nwSUSY}
(supersymmetric models), \cite{nwextHS} (non-supersymmetric
extended Higgs sectors), \cite{nwcomposite} (composite models),
\cite{nwextradim} (models with extra dimensions),
\cite{nwleptoquarks} (leptoquarks), and \cite{nwothers}
(other models).}
Since the advertised discrepancy comes from the calculation of
Ref.~\cite{Davier:1998si} of the hadronic vacuum polarization
contributions $a_{\mu}^{\rm (v.p.)}$ to $a_{\mu}$, 
we re--evaluate in the present paper the theoretical (pQCD+OPE) parts 
of this quantity. We use the same theoretical approach as in
Ref.~\cite{Davier:1998si}, which in turn is based on the approach
of Ref.~\cite{Groote:1998pk}. However, in addition, we take into 
account the known renormalon structure of the Adler function.

\section{Formalism}

\begin{figure}[ht]
 \centering\epsfig{file=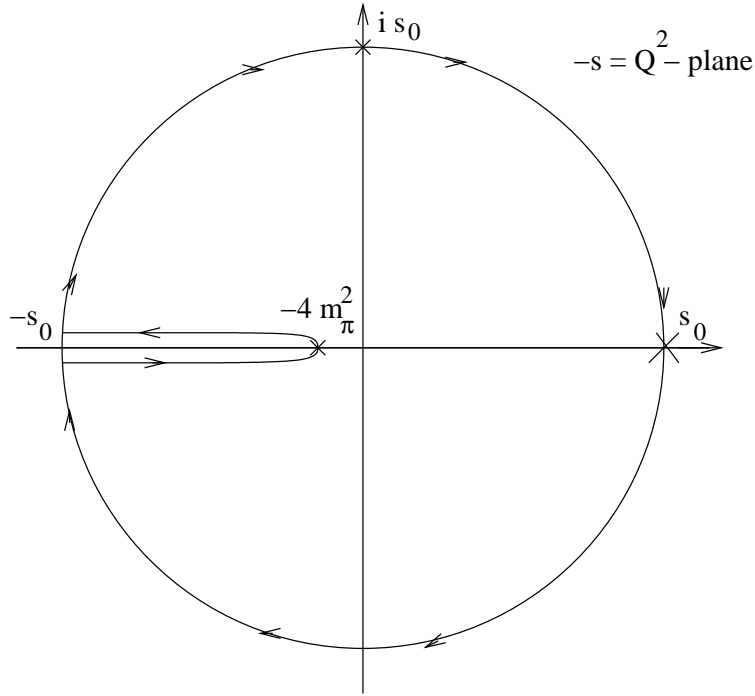}
\vspace{0.3cm}
\caption{\footnotesize
Application of the Cauchy theorem to the depicted
integration contour in the $Q^2$--plane leads to the
contour integral in Eq.~(\ref{dec2}).
}
\label{fig1}
\end{figure}

According to Ref.~\cite{Brodsky:1968sr}, the hadronic vacuum 
polarization contribution to the muon anomalous
magnetic moment $a_{\mu} \equiv (g-2)/2$ appears in the following
dispersion integral:
\begin{equation}
a_{\mu}^{\rm (v.p.)} = \frac{\alpha_{em}^2(0)}{3 \pi^2}
\int_{4 m^2_{\pi}}^{\infty} \frac{ds}{s} K(s) R_{e^+e^-}(s) \ ,
\label{di}
\end{equation}
where  $K(s)$ is the QED
kernel \cite{Brodsky:1968sr}
\begin{eqnarray}
K(s) &=& x^2 \left( 1 - \frac{x^2}{2} \right) +
(1 + x)^2 \left( 1 + \frac{x^2}{2} \right)
\left[ \ln(1\!+\!x) - x + \frac{x^2}{2} \right]
+ \frac{(1+x)}{(1-x)} x^2 \ln x \ .
\label{Ks}
\end{eqnarray}
Here, $x=(1 - y_{\mu})/(1 + y_{\mu})$ with
$y_{\mu} = (1 - 4 m^2_{\mu}/s)^{1/2}$.
The largest part (about $92 \%$) of $a_{\mu}^{\rm (v.p.)}$
comes from the region with CMS energy $\sqrt{s} < \sqrt{s_0} = 1.8$ GeV.
Following the approach of Ref.~\cite{Groote:1998pk}, applied in
Ref.~\cite{Davier:1998si} to $a_{\mu}^{\rm (v.p.)}$, we rewrite
the dispersion integral (\ref{di}) (with $s_{\rm max} = s_0$)
in the form
\begin{eqnarray}
\frac{3 \pi^2}{\alpha_{em}^2(0)} \times
a_{\mu}^{\rm (v.p.)}(s\leq s_0) &=&
\int_{4 m^2_{\pi}}^{s_0} ds \; R_{e^+e^-}(s)
\left[ \frac{K(s)}{s}  - C_1 \left( 1 - \frac{s}{s_0} \right) \right]
\nonumber\\
&& + \; C_1 \int_{4 m^2_{\pi}}^{s_0} ds \; R_{e^+e^-}(s)
\left( 1 - \frac{s}{s_0} \right) \ .
\label{dec1}
\end{eqnarray}
Here $C_1$ is in principle an arbitrary constant,
which however, according to the philosophy of Ref.~\cite{Groote:1998pk},
may be chosen in such a way as to minimize the first
(``data'') integral and maximize the second (``theory'')
integral. It is known that $R_{e^+e^-}(s) =
12 \pi {\rm Im} \Pi(s + {\rm i} \varepsilon)$, with
$\Pi(s)$ being the hadronic part of the (vector) photon vacuum
polarization function which has no poles in the interval
$[0, 4 m^2_{\pi})$; further, $(1 - s/s_0)$ also has no poles
in that interval, in contrast to the function $K(s)$.
Therefore, the Cauchy theorem can be applied to the
second (``theory'') integral, with the path of Fig.~\ref{fig1}.
Carrying subsequently integration by parts, and
using the identity $D(Q^2\!\equiv\!-s) = - 12 \pi^2 s \; d \Pi(s)/ds$,
leads to
\begin{eqnarray}
\frac{3 \pi^2}{\alpha_{em}^2(0)} \times
a_{\mu}^{\rm (v.p.)}(s\leq s_0) & = &
\int_{4 m^2_{\pi}}^{s_0} ds \; R_{e^+e^-}(s)
\left[ \frac{K(s)}{s}  - C_1 \left( 1 - \frac{s}{s_0} \right) \right]
\nonumber\\
&& + \frac{C_1 s_0}{4 \pi} \int_{-\pi}^{\pi} dy \;
D(Q^2\!=\!s_0 {\rm e}^{{\rm i} y}) (1 + {\rm e}^{{\rm i} y})^2 \ ,
\label{dec2}
\end{eqnarray}
where the associated vector Adler function $D(Q^2)$
can be written in the following way:\footnote{
This Adler function does not contain QED radiative corrections.
The effects of them will be added later on.}
\begin{equation}
D(Q^2)  =  N_{\rm c} \sum_{f=u,d,s} Q_f^2
\left[ 1 + D_{\rm can.}(Q^2) + 4 \pi^2 \times \sum_{d=2,4,\ldots}
D_{f {\bar f}; V}^{(d;J\!=\!1)}(Q^2) \right] \ .
\label{D1}
\end{equation}
Here, $D_{\rm can.}(Q^2)$ is the canonically normalized
massless QCD part with dimension $d=0$, whose power expansion in
$a^{\overline {\rm MS}}(Q^2) \equiv \alpha_s^{\overline {\rm MS}}(Q^2)/\pi$
is
\begin{equation}
D_{\rm can.}(Q^2) = a^{\overline {\rm MS}}(Q^2) \times
\left[ 1 + d_1^{(0)} a^{\overline {\rm MS}}(Q^2) +
d_2^{(0)} \left( a^{\overline {\rm MS}}(Q^2) \right)^2 +
d_3^{(0)} \left( a^{\overline {\rm MS}}(Q^2) \right)^3 + \cdots
\right] \ ,
\label{Dcan}
\end{equation}
with $d_1^{(0)} = 1.6398$ \cite{coeffs1},
$d_2^{(0)} = 6.3710$ \cite{coeffs2},
and $d_3^{(0)}$ is estimated to be $d_3^{(0)} = 25 \pm 10$
\cite{Cvetic:2001sn}. The renormalization scale in
(\ref{Dcan}) is $\mu^2\!=\!Q^2$, and the renormalization
scheme is ${\overline {\rm MS}}$.
The number of active quark flavors
is $n_f\!=\!3$.
The $d=2$ contributions are \cite{Braaten:1992qm,PP}
\begin{eqnarray}
4 \pi^2 D_{f {\bar f}; V}^{(d = 2;J = 1)}(Q^2)
&=& - \frac{6 m_f^2(Q^2)}{Q^2} \left[ 1 +
\frac{14}{3} a^{\overline {\rm MS}}(Q^2)
+ 43.0581 \left(a^{\overline {\rm MS}}(Q^2)\right)^2
+ {\cal O}(a^3) \right] \ ,
\label{Dd=2}
\end{eqnarray}
where only the $s$ quark contributes appreciably.
The $d=4$ contributions are those of the gluon condensate
\cite{Braaten:1992qm}
\begin{equation}
4 \pi^2 D_{f {\bar f}; V}^{(\rm glc; d = 4;J = 1)}(Q^2)
= + \frac{2 \pi^2}{3 (Q^2)^2} \langle a GG \rangle
\left[ 1 - \frac{11}{18}
a^{\overline {\rm MS}}(Q^2) + {\cal O}(a^2) \right] \ ,
\label{Dd=4glc}
\end{equation}
those of the quark mass condensates \cite{Braaten:1992qm}
\begin{eqnarray}
\lefteqn{
4 \pi^2 D_{f {\bar f}; V}^{(\rm qc.; d=4;J=1)}(Q^2)
=  \frac{16 \pi^2}{(Q^2)^2} \left[
1 + \frac{1}{3} a^{\overline {\rm MS}}(Q^2) +
\frac{11}{2} \left(a^{\overline {\rm MS}}(Q^2)\right)^2
+ {\cal O}(a^3) \right] \langle m_f {\bar f} f \rangle \
}
\nonumber\\
&& + \frac{8 \pi^2}{(Q^2)^2} \left[
\frac{4}{27} a^{\overline {\rm MS}}(Q^2) +
1.074 \left(a^{\overline {\rm MS}}(Q^2)\right)^2
+ {\cal O}(a^3) \right]
\sum_{f_k=u,d,s} \langle m_{f_k} {\bar f_k} f_k \rangle \  ,
\label{Dd=4qc}
\end{eqnarray}
and those proportional to $m_f^4$ \cite{Braaten:1992qm}
\begin{eqnarray}
4 \pi^2 D_{f {\bar f}; V}^{(\rm qm; d=4;J=1)}(Q^2)
& = & \frac{48}{7 (Q^2)^2} \left[
- \frac{1}{a^{\overline {\rm MS}}(Q^2)} + 1 + 12.0
a^{\overline {\rm MS}}(Q^2)
+ {\cal O}(a^2) \right] m^4_f(Q^2)
\nonumber\\
&& - \frac{2}{7 (Q^2)^2} \left[ 1 + 8.4 a^{\overline {\rm MS}}(Q^2)
+ {\cal O}(a^2) \right]
\sum_{f_k=u,d,s} m^4_{f_k}(Q^2) \ .
\label{Dd=4qm}
\end{eqnarray}
The terms with dimension $d \geq 6$ do not
contribute to the ``theory'' part in (\ref{dec2}) in the leading order
renormalization group (RG) approximation.
We note that the leading term
in (\ref{Dd=4qc}) is twice as large as that in \cite{Davier:1998si}
[their Eq.~(9)]. We will see
later that the terms (\ref{Dd=4qm}) give negligible contributions,
but not the terms (\ref{Dd=2})--(\ref{Dd=4qc}).
In the quark condensate terms (\ref{Dd=4qc}) we can use
the (approximate) leading PCAC relations 
$m_s \langle {\bar q} q \rangle \approx - f^2_{\pi} (m^2_{K}\!-\!m^2_{\pi})$
and $(m_u\!+\!m_d) \langle {\bar q} q \rangle \approx - f^2_{\pi} m^2_{\pi}$ 
($q\!=\!u,d$ or $s$; $f_{\pi}\!=\!0.0924 \pm 0.0003$ GeV)
\begin{eqnarray}
\langle A_{\rm qc}^{(1)} \rangle &\equiv&
\sum_{f=u,d,s} Q_f^2 \langle m_f {\bar f} f \rangle
\approx - 2.50 \times 10^{-4} \ {\rm GeV}^4 \ ,
\label{qc1}
\\
\langle A_{\rm qc}^{(2)} \rangle &\equiv&
\sum_{f=u,d,s} \langle m_f {\bar f} f \rangle
\approx - 2.08 \times 10^{-3} \ {\rm GeV}^4 \ .
\label{qc2}
\end{eqnarray}
Within the leading PCAC approach, the uncertainties in the numbers 
in (\ref{qc1})--(\ref{qc2}) originate from the uncertainty of
$f^2_{\pi}$, and of the ratio
$\epsilon_u\!\equiv\!m_u/m_s = 0.029 \pm 0.003$
\cite{Leutwyler:1996eq,PP}. They are $\sim\!1 \%$,
and affect insignificanly the results of the present paper. 

Insertion of the expressions (\ref{Dcan})--(\ref{Dd=4qm})
into the contour integral (\ref{dec2}) gives
\begin{eqnarray}
\lefteqn{
\frac{3 \pi^2}{\alpha_{em}^2(0)} \times
a_{\mu}^{\rm (v.p.)}(s\leq s_0)^{\rm (theor.)} =
C_1 s_0 \times \Bigg\{ 1 + A_{\rm can.}
}
\nonumber\\
&& - \frac{1}{2 \pi} \frac{m_s^2(s_0)}{s_0}
M_{0,2} \left[ 1 + \frac{14}{3} \frac{M_{1,2}}{M_{0,2}}
+ 43.0581 \frac{M_{2,2}}{M_{0,2}} \right]
+ \frac{\pi}{3} \frac{ \langle a GG \rangle }{s_0^2}
A_{0,4} \left[ 1 - \frac{11}{18} \frac{A_{1,4}}{A_{0,4}} \right]
\nonumber\\
&& + 12 \pi \frac{1}{s_0^2} \langle A_{\rm qc}^{(1)} \rangle A_{0,4}
\left[1 + \frac{1}{3}\frac{A_{1,4}}{A_{0,4}} +
\frac{11}{2} \frac{A_{2,4}}{A_{0,4}} \right]
+ 4 \pi \frac{1}{s_0^2} \langle A_{\rm qc}^{(2)} \rangle \frac{4}{27}
A_{1,4} \left[1 + 7.25 \frac{A_{2,4}}{A_{1,4}} \right]
L\nonumber\\
&& - \frac{4}{7 \pi} \frac{m_s^4(s_0)}{s_0^2}
M_{-1,4} \left[ 1 -
\frac{M_{0,4}}{M_{-1,4}} - 12. \frac{M_{1,4}}{M_{-1,4}} \right]
- \frac{1}{7 \pi} \frac{m_s^4(s_0)}{s_0^2}
M_{0,4} \left[ 1 + 8.4
\frac{M_{1,4}}{M_{0,4}} \right]
\Bigg\} \ .
\label{series}
\end{eqnarray}
Here, we used the complex momentum contour integrals
\begin{eqnarray}
A_{\rm can.} &=& \frac{1}{2 \pi} \int_{-\pi}^{\pi}
dy (1 + {\rm e}^{ {\rm i} y} )^2 D_{\rm can.}(
Q^2\!=\!s_0 {\rm e}^{ {\rm i} y}) \ ,
\label{acancont}
\\
A_{n,2k} & = & \int_{-\pi}^{\pi} dy
(1 + {\rm e}^{ {\rm i} y} )^2 {\rm e}^{-{\rm i} k y}
\left( a^{\overline {\rm MS}}(Q^2\!=\!s_0{\rm e}^{ {\rm i} y}) \right)^n \ ,
\label{Acont}
\\
M_{n,2k} & = & \int_{-\pi}^{\pi} dy
(1 + {\rm e}^{ {\rm i} y} )^2 {\rm e}^{-{\rm i} k y}
\left( \frac{ m_s(s_0{\rm e}^{ {\rm i} y})}{m_s(s_0)} \right)^{2 k}
\left( a^{\overline {\rm MS}}(Q^2\!=\!s_0{\rm e}^{ {\rm i} y}) \right)^n \ .
\label{Mcont}
\end{eqnarray}
For RGE evolution of $a^{\overline {\rm MS}}(Q^2)$ we use the
four--loop truncated perturbation
series (TPS) \cite{vanRitbergen:1997va}
of the ${\overline {\rm MS}}$ beta function, with $n_f\!=\!3$.
In addition, for the RGE evolution of $m_s(Q^2)$ we use
the ${\overline {\rm MS}}$ four--loop
TPS quark mass anomalous dimension \cite{Vermaseren:1997fq}.
For example, the RGE evolution along the complex
momentum contour gives
\begin{eqnarray}
\frac{ m_s(s_0{\rm e}^{ {\rm i} y})}{m_s(s_0)}
& = & \exp \Bigg\{ - {\rm i}
\int_0^y d y' a^{\overline {\rm MS}}(s_0 {\rm e}^{{\rm i} y'})
\left[ 1 + \sum_{n=1}^3 {\widetilde \gamma}_n
\left( a^{\overline {\rm MS}}(s_0 {\rm e}^{{\rm i} y'}) \right)^n
\right] \Bigg\} \ ,
\label{msevol}
\end{eqnarray}
with ${\widetilde \gamma}_1 = 3.79167$,
${\widetilde \gamma}_2 = 12.4202$, and
${\widetilde \gamma}_3 = 44.263$ \cite{Vermaseren:1997fq}.

\section{Evaluation}

We first use the input values as used in Ref.~\cite{Davier:1998si}
\begin{eqnarray}
\langle a GG \rangle & = & (0.015 \pm 0.020) \ {\rm GeV}^4,
\label{aGG1}
\\
\alpha_s^{\overline {\rm MS}}(m^2_{\tau}) &=& 0.333 \pm 0.017
\quad \left( \Rightarrow \
\alpha_s^{\overline {\rm MS}}(M^2_{\rm z}) \approx 0.1201 \pm 0.0020
\right) \ ,
\label{alphas1}
\\
m_s(1 {\rm GeV}^2) &=& 0.20 \pm 0.07 \ {\rm GeV}
\quad \Rightarrow \ m_s(m^2_{\tau}) \approx 0.151 \pm 0.053 \ {\rm GeV} \ .
\label{ms1G1}
\end{eqnarray}
Further, we use for $D_{\rm can.}(Q^2)$ the NNLO TPS
(i.e., with $d_3^{(0)}=0$) and with the renormalization scale
$\mu^2=Q^2$, i.e., the approach apparently used by
\cite{Davier:1998si}. The result for
their input values, and for $C_1=0.007 \ {\rm GeV}^{-2}$, is then:
\begin{equation}
10^{11} \times a_{\mu}^{\rm (v.p.)}(s \leq s_0; {\rm th. \  part};
{\rm NNLO} \ D_{\rm can.})
= 4752 \pm 108 \ ,
\label{amuNNLO}
\end{equation}
in contrast to their value $(4686.2 \pm 113.2)$. The central
values of these theory parts are thus higher by $1.4 \%$
than those given in \cite{Davier:1998si}. This percentage does not change
at different values of the parameter $C_1$, since the results are linearly
proportional to $C_1$.
The uncertainty $\pm 108$ in (\ref{amuNNLO}) is obtained by
adding in quadrature the uncertainty from $\alpha_s$
($\pm 61$), from $m_s$ ($\pm 74$), and from
$\langle a GG \rangle$ ($\pm 49$).
The separate contributions to the central value $4752$ in
(\ref{amuNNLO}) are:
$4079$ from the leading term;
$749$ from the (resummed) canonical part (\ref{Dcan})
[ $\Rightarrow$ (\ref{acancont})];
$-87$ from the $d\!=\!2$ strange mass term (\ref{Dd=2});
$37$ from the $d\!=\!4$ gluon condensate term (\ref{Dd=4glc});
$-25$ from the $d\!=\!4$ quark condensate terms (\ref{Dd=4qc});
$-1$ from the $d\!=\!4$ quark mass terms (\ref{Dd=4qm}).

We can, however, re--calculate the canonical part (\ref{acancont})
by using methods which account for the renormalon structure
of the Adler function. The latter structure has been shown
\cite{Cvetic:2001sn,Cvetic:2001ws} to have numerically significant 
effects in the hadronic $\tau$ decay width ratio.\footnote{
In Refs.~\cite{Cvetic:2001sn,Cvetic:2001ws}, the massless
QCD quantity $r_{\tau}$ was calculated (resummed).
It is obtained from the hadronic $\tau$ decay width
ratio $R_{\tau}$ by subtracting from it the strangeness--changing
contributions, factoring out the CKM--element
$|V_{ud}|^2$ and the electroweak correction factors,
and subtracting the quark mass ($m_{u,d}\not=0$) contributions.
The quantity $r_{\tau}$ is the same kind of
contour integral in the complex momentum plane as the quantity
$A_{\rm can.}$ of Eq.~(\ref{acancont}), but with the contour
factor $(1 + {\rm e}^{{\rm i} y})^2$ replaced by
$(1 + {\rm e}^{{\rm i} y})^3 (1 - {\rm e}^{{\rm i} y})$.
Therefore, the analyses of Refs.~\cite{Cvetic:2001sn,Cvetic:2001ws}
can be repeated, without any major changes, for the resummation of
the quantity $A_{\rm can.}$ of Eq.~(\ref{acancont}).}
Ref.~\cite{Cvetic:2001sn} uses ordinary Borel transforms,
and Ref.~\cite{Cvetic:2001ws} modified Borel transforms of
$D_{\rm can.}$. In these two References, the exactly known 
leading infrared renormalon singularity of the Borel transform
$\widetilde{D}_{\rm can.}(b)$, at the value of the Borel 
variable $b\!=\!2$, has been fully taken into account
as the pre--factor function of the form $1/(1-b/2)^{1+\nu}$ 
and $1/(1-b/2)$, respectively. This procedure allows us to describe
the Borel transform more accurately in the most important region
of the Borel integration, namely, the interval between the
origin and the leading IR renormalon at $b=2$.
Furthermore, judiciously chosen conformal transformations
$b\!=\!b(w)$ have been employed there
to map the singularities of other renormalons
further away from the origin and thus to reduce their
numerical significance. For details of the calculation procedures, 
we refer to these two References. 

If we re--calculate the caconical part (\ref{acancont})
by using the ordinary Borel transform method of 
Ref.~\cite{Cvetic:2001sn}, the
result (\ref{amuNNLO}) increases further to
$4817 \pm 116$, as a consequence of the renormalon effects.
In $D_{\rm can.}$ we took $d_3^{(0)} = 25$, and for the
renormalization scale (RScl) in $D_{\rm can.}$ we took the value
which gives us the local insensitivity of the result with
respect to RScl ($\mu^2 \approx 1.6 Q^2$), as argued in
\cite{Cvetic:2001sn}.

If we apply to expression (\ref{acancont})
the method of resummation of Ref.~\cite{Cvetic:2001ws}
which employs modified Borel transforms, with the same input
and with $d_3^{(0)} = 25\pm 10$,\footnote{
We refer to \cite{Cvetic:2001sn} for a detailed discussion
of this estimate of $d_3^{(0)}$ values.} 
we obtain a value very similar to the aforementioned one
\begin{equation}
10^{11} \times a_{\mu}^{\rm (v.p.)}(s \leq s_0; {\rm th. \ part};
{\rm meth. \cite{Cvetic:2001ws}})
= 4820 \pm 120 \ .
\label{amumBT1}
\end{equation}
The uncertainty $\pm 120$ in (\ref{amumBT1}) is obtained by
adding in quadrature the uncertainty from $\alpha_s$
($\pm 74$), from $m_s$ ($\pm 75$), from
$\langle a GG \rangle$ ($\pm 49$), as well as from
the resummation method uncertainty
and the uncertainty of $d_3^{(0)}$ ($\pm 28$).
This calculation and the uncertainty estimates procedure 
are carried out in complete analogy
with Ref.~\cite{Cvetic:2001ws}, to which we refer for details.
The separate contributions to the central value $4820$ in (\ref{amumBT1})
are the same as those to (\ref{amuNNLO}), except that
the contribution from the (resummed) canonical part (\ref{Dcan})
is now $817$ (before: $749$).

In order to isolate the contribution of the renormalon structure
included in the value (\ref{amumBT1}), we should compare
the latter central value with the one obtained by using for 
$D_{\rm can}$ the ${\rm N}^3 {\rm LO}$ TPS with $d_3^{(0)}\!=\!25$.
The central value result in this case is $4772$. The obtained 
renormalon structure effect is thus $48$ units, or $1.0 \%$.  

The input values (\ref{alphas1})--(\ref{ms1G1}), used by
the authors of Ref.~\cite{Davier:1998si}, and taken
up until now in the present work, can be replaced
by what we believe to be more updated values
\begin{eqnarray}
\alpha_s^{\overline {\rm MS}}(m^2_{\tau}) &=& 0.3254 \pm 0.0124
\quad \left( \Rightarrow \
\alpha_s^{\overline {\rm MS}}(M^2_{\rm z}) \approx 0.1192 \pm 0.0015
\right) \ ,
\label{alphas2}
\\
m_s(m^2_{\tau}) &=& 0.119 \pm 0.024 \ {\rm GeV} \ .
\label{msmtau}
\end{eqnarray}
The values (\ref{alphas2}) were obtained in \cite{Cvetic:2001ws}
by a detailed analysis of the $R_{\tau}$ ratio, involving
modified Borel transforms, and accounting for the renormalon
structure of the associated Adler function via an explicit ansatz
and with judiciously chosen conformal transformations.
This result virtually agrees with the one obtained
in the $R_{\tau}$--analysis  of Ref.~\cite{Cvetic:2001sn}
where ordinary Borel transforms were used instead.
The values (\ref{alphas2}) are shifted downwards and the uncertainties
are reduced, in comparison to the values (\ref{alphas1}).
The latter values are based largely on the ALEPH analysis
of the $R_{\tau}$ decay \cite{Barate:1998uf}. The latter
analysis did not account for the renormalon structure of the
associated Adler function.
The trend towards smaller values of $\alpha_s$ and
towards smaller uncertainties appears also in the
analysis of the $R_{\tau}$ ratio of the authors of
Ref.~\cite{Maxwell:2001uv},
who accounted for the renormalon structure via a
large--$\beta_0$ resummation of the ordinary Borel transform
and employed a resummation related to the effective charge (ECH)
method -- they obtained $\alpha_s(m^2_{\tau}) = 0.330 \pm 0.014$.
The values (\ref{msmtau}) for the strange quark mass, which
are significantly lower than those in (\ref{ms1G1}),
were obtained in the recent analysis of Ref.~\cite{PP}.

Further, ALEPH analysis \cite{Barate:1998uf}
of the $\tau$ decays predicts the gluon condensate
term to be consistent with zero
\begin{equation}
\langle a GG \rangle =  (0.001 \pm 0.015)  \ {\rm GeV}^4 \ ,
\label{aGG2}
\end{equation}
in contrast with the input (\ref{aGG1}).
We will take, in addition to the $(\alpha_s, m_s)$--inputs 
(\ref{alphas2})--(\ref{msmtau}),
either the input (\ref{aGG1}) or (\ref{aGG2}) for the
gluon condensate term.
Small values of the gluon condensate
close to the ALEPH values (\ref{aGG2}) are also suggested in
the formalism of Ref.~\cite{Lee:2001ws}, where the power--suppressed
terms are obtained from the knowledge of the perturbation
series of $D_{\rm can}$ (\ref{Dcan}) and of its infrared
renormalon structure.

Applying then again the resummation method
of Ref.~\cite{Cvetic:2001ws} to expression (\ref{acancont}), 
we obtain the prediction
\begin{eqnarray}
10^{11} \times a_{\mu}^{\rm (v.p.)}(s \leq s_0; {\rm th. \ part};
{\rm meth. \cite{Cvetic:2001ws}})
&=& 4823 \pm 77 \quad
{\rm for} \ \langle aGG \rangle \ {\rm value \ Eq.~(\ref{aGG1})}
\label{amumBT2a}
\\
&=& 4789 \pm 71  \quad
{\rm for} \ \langle aGG \rangle \ {\rm value \ Eq.~(\ref{aGG2})}
\ .
\label{amumBT2b}
\end{eqnarray}
The uncertainties $\pm 77, \pm 71$ in (\ref{amumBT2a})--(\ref{amumBT2b})
are obtained by adding in quadrature the uncertainty from $\alpha_s$
($\pm 49$), from $m_s$ ($\pm 24$), from
$\langle a GG \rangle$ ($\pm 49, \pm 37$), and from the resummation
method uncertainty and $d_3^{(0)}$--uncertainty ($\pm 25$).
The separate contributions to the central value $4823$ in
(\ref{amumBT2a}) are:
$4079$ from the leading term;
$787$ from the (resummed) canonical part (\ref{Dcan});
$-54$ from the $d\!=\!2$ strange mass term (\ref{Dd=2});
$37$ [$2$ if using (\ref{aGG2})]
from the $d\!=\!4$ gluon condensate term (\ref{Dd=4glc});
$-25$ from the $d\!=\!4$ quark condensate terms (\ref{Dd=4qc});
$-0.5$ from the $d\!=\!4$ quark mass terms (\ref{Dd=4qm}).

We note that the uncertainties $\pm 77, \pm 71$ as obtained by
us in (\ref{amumBT2a})--(\ref{amumBT2b})
are lower than the
uncertainty $\pm 113.2$ for that quantity obtained by
the authors of Ref.~\cite{Davier:1998si} 
(for $C_1=0.007 \ {\rm GeV}^{-2}$).
The main reason for this is that we used different,
in our opinion more updated, values of
$\alpha_s^{\overline {\rm MS}}(m^2_{\tau})$ and
$m_s(m^2_{\tau})$ (\ref{alphas2})--(\ref{msmtau})
than the values (\ref{alphas1})--(\ref{ms1G1}) used by
Ref.~\cite{Davier:1998si}.
The uncertainties $\pm 77, \pm 71$ are about the same as the uncertainties
obtained in Ref.~\cite{Davier:1998si} for the ``data'' part,
i.e., the first integral in (\ref{dec2}), for 
$C_1=0.000$--$0.007 \ {\rm GeV}^{-2}$
(see Table.~1 of \cite{Davier:1998si}). The authors of
Ref.~\cite{Davier:1998si} chose 
$C_1 \approx 0.001$--$0.002 \ {\rm GeV}^{-2}$,
i.e., their ``theory'' contribution was very small
in comparison to their ``data'' contribution. The argument for
the virtual exclusion of the theory from their considerations
was that the uncertainties from their ``theory'' part were
too high when $C_1$ is appreciable. We believe that
a different approach, which emphasizes the ``theory''
part more than the ``data'' part, is legitimate as well,
especially because the uncertainties from our analysis
of the ``theory'' part are reasonable and comparable to
the uncertainties of the ``data'' part even for large
values of $C_1$. 
The authors of Ref.~\cite{Davier:1998si} obtained the values of 
the ``data'' parts for $C_1 \leq 0.007 \ {\rm GeV}^{-2}$, 
by using the available $e^+e^-$ and $\tau$ decay data.
Therefore, we choose the largest $C_1=0.007 \ {\rm GeV}^{-2}$ 
listed in their Table 1. 

The ``data'' part given in their Table 1,
for $C_1=0.007 \ {\rm GeV}^{-2}$, is
\begin{equation}
10^{11} \times a_{\mu}^{\rm (v.p.)}(s \leq s_0; {\rm data \ part})
= 1622 \pm 56 \ .
\label{amudata}
\end{equation}
It corresponds to the first integral in Eq.~(\ref{dec2}).

\begin{figure}[ht]
 \centering\epsfig{file=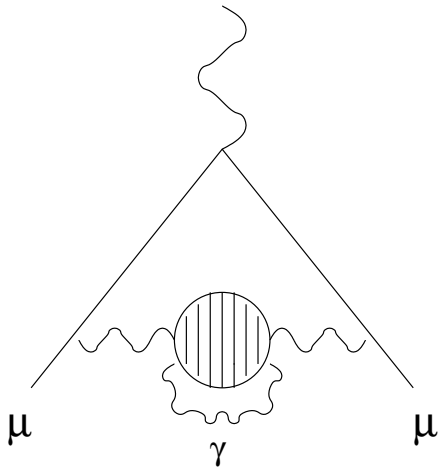}
\vspace{0.3cm}
\caption{\footnotesize
QED correction to the hadronic blob.
}
\label{fig2}
\end{figure}

Our evaluation of the theoretical part
(\ref{amumBT2a})--(\ref{amumBT2b}) of $s \leq s_0$ contributions
excludes any QED corrections to the hadronic vacuum polarization.
The leading term of such QED corrections corresponds 
to the exchange of a virtual
photon within the hadronic blob in Fig.~\ref{fig2}. This diagram
induces in $R_{e^+e^-}$ in the master formula (\ref{di}) 
the real photon emissions and the virtual
(plus soft photon emissions) corrections to the pure hadronic final
states. Since
the theory part in Eq.~(\ref{dec2}) (2nd line) is
dominated by the perturbative QCD (pQCD) Adler function,
this QED correction can be easily implemented by replacing
$(1 + D_{\rm can.}(Q^2))$ in (\ref{D1}) with the QED corrected one:
\bear
\left(1 + D_{\rm can.}(Q^2)\right)\left(1+ \frac{3 Q_f^2}{4\pi}
\alpha_{{\rm em}}(Q^2)\right)\,.
\label{qedfactor}
\eear
The QED corrections to the power-suppressed terms in (\ref{D1})
are negligibly small.
Then, since the running of $\alpha_{{\rm em}}(Q^2)$ along the integration
contour can be
safely ignored,  the QED correction can be seen to shift the
theory part by the factor:
\bear
\left[1+ \frac{3}{4\pi}(\sum Q_f^4)/(\sum Q_f^2)  
\alpha_{{\rm em}}(s_0)\right]=
1+ \alpha_{{\rm em}}(s_0)/4\pi
\approx 1+6\times 10^{-4}
\eear
Thus, the QED correction to the theory part of $a_\mu^{\rm (v.p)}$
is approximately $(6\times 10^{-4})\times 4800 \approx 3$.
It is small.

On the other hand, in the data part (\ref{amudata}),
the aforementioned class of QED contributions is, 
in principle, already included.
This data part is taken from Ref.~\cite{Davier:1998si},
and corresponds to the first integral in Eq.~(\ref{dec2}),
for $C_1\!=\!0.007 \ {\rm GeV}^{-2}$, with $R_{e^+e^-}(s)$ there
based largely on the experimental $e^+ e^-$ and $\tau$--decay data. 
In this connection, we mention that there is some
controversy on whether the specific photon
decay channels via  $\rho$ meson 
[$e^+e^- \to (\rho) \to \pi \pi \gamma$] are
included in the data or should be added to them.
The contribution of these channels to the
data part (\ref{amudata}), in the narrow width approximation,
would be about $(18 \pm 4)$.
However, the $\tau$--decay data already include this
decay channel, while the situation with the
$e^+ e^-$ data in this respect is less clear.
Since the authors of Ref.~\cite{Davier:1998si}
included in their analysis both types of data,
we presume that the additional possible contributions 
from the aforementioned photonic decay channels to the data part
(\ref{amudata}) are significantly lower than
$(18 \pm 4)$, and we will therefore ignore such
contributions.

We note that the inclusive, total QED correction  to $a_\mu^{\rm (v.p.)}$ is
very small when compared to the contribution from the
real photon emissions. From the above QED correction to the
theory part we can estimate the total QED correction to $a_\mu^{\rm (v.p.)}$
as $a_\mu^{\rm (v.p.)}\times \alpha_{\rm em}(s_0)/4\pi\approx 7000\times
(6\times 10^{-4})\approx 4$.
This is much smaller than the contribution from the
real photon emissions, which is according to \cite{TY1TY2} about 90 units
from the photonic decay channels of mesons, e.g.
$\rho$, $\omega$, and $\phi$.
These two contributions may appear incompatible, but there is no real
contradiction.
It can be seen most easily in the narrow width approximation (n.w.a.)
of the resonances. The n.w.a. formulas for the portions of $R_{e+e-}$
in (\ref{di}) from the various decay channels of a resonance
contain as factors the corresponding branching ratios, whose sum is always one.
Thus, an increase in the photonic decay channel contributions must be
compensated  by the decrease in the 
pure hadronic channel contributions by exactly the same amount.
The contributing QED correction factor in the Adler function in 
Eq.~(\ref{qedfactor}) thus accounts for the  QED corrections in the continuum
only.
Beyond the n.w.a. approximation, however, the cancellation between the photonic
channel contributions and the pure hadronic channel contributions
cannot be exact, since the decay widths of the resonances are sizable, and
so a small QED correction survives after the cancellation.
This remaining QED correction,
along with the QED corrections in the continuum, are those accounted for by
the QED correction factor in the Adler function in Eq.~(\ref{qedfactor}).

%It was derived in the partonic picture, is thus independent of the details 
%of the resonances and hence remains valid beyond the n.w.a. framework.
%Of course, in real world, this cancellation between the photon
%channel contributions and the pure hadronic channel contributions
%cannot be exact, since the decay widths of the resonances are sizable, and
%so a small QED correction survives after the cancellation.
%This remaining QED correction,
%along with other QED corrections in the continuum, are those accounted for by
%the QED correction factor in the Adler function in Eq.~(\ref{qedfactor}).

We now include in our theory--part contributions
(\ref{amumBT2a})--(\ref{amumBT2b}) 
the aforementioned QED contributions (\ref{qedfactor}).
\begin{eqnarray}
10^{11} \times a_{\mu}^{\rm (v.p.)}(s \leq s_0; {\rm th. \ part};
{\rm meth. \cite{Cvetic:2001ws}})
&=& 4826 \pm 77 \quad
{\rm for} \ \langle aGG \rangle \ {\rm value \ Eq.~(\ref{aGG1})}
\label{amumBT3a}
\\
&=& 4792 \pm 71  \quad
{\rm for} \ \langle aGG \rangle \ {\rm value \ Eq.~(\ref{aGG2})}
\ .
\label{amumBT3b}
\end{eqnarray}
This implies for the sum of the ``theory''
(\ref{amumBT3a})--(\ref{amumBT3b})
and the ``data'' (\ref{amudata}) parts
\begin{eqnarray}
10^{11} \times a_{\mu}^{\rm (v.p.)}(s \leq s_0)
&=& 6448 \pm 95 \quad
{\rm for} \ \langle aGG \rangle \ {\rm value \ Eq.~(\ref{aGG1})}
\label{amuthdata}
\\
&=& 6414 \pm 90 \quad
{\rm for} \ \langle aGG \rangle \ {\rm value \ Eq.~(\ref{aGG2})}
\ .
\label{amuthdatb}
\end{eqnarray}
The authors of Ref.\cite{Davier:1998si} obtained for this
quantity the value $6343 \pm 60$, by almost excluding
their ``theory'' part (complete exclusion of the ``theory''
part gave them $6350.5 \pm 74$).

If we decreased parameter $C_1$ from $0.007 \ {\rm GeV}^{-2}$
toward zero, our central values in (\ref{amuthdata})--(\ref{amuthdatb})
would go down toward the pure ``data'' value $6351$, i.e.,
by $\pm 97$ and $\pm 63$ units, respectively. We will comment
on this point in the next Section.
 
If we add to the obtained quantities (\ref{amuthdata})--(\ref{amuthdatb})
the  hadronic vacuum polarization parts from the region $s > s_0$
as obtained in Ref.~\cite{Davier:1998si},\footnote{
They obtained $581 \pm 15$ for this contribution,
using for the pQCD parts apparently the value
(\ref{alphas1}) for $\alpha_s$. The authors of
Ref.~\cite{TY1TY2} obtained slightly higher values
$584 \pm 9$, once we subtract from their values
the (pQCD) contributions from
$(3 \ {\rm GeV}^2 < s < s_0)$ ($s_0\!=\!1.8^2 \ {\rm GeV}^2$).
They used lower values for $\alpha_s$:
$\alpha_s(m^2_{\tau}) \approx 0.3088 \pm 0.0245$.
This small discrepancy would apparently increase
once we adjusted $\alpha_s$ to the same value,
say (\ref{alphas1}), for the two approaches
of Refs.~\cite{Davier:1998si,TY1TY2}). Then the
prediction of Ref.~\cite{TY1TY2} would be about $592 \pm 10$.
We decide to take the value of Ref.~\cite{Davier:1998si}: $581 \pm 15$.
We can reproduce their pQCD--parts with the simple TPS
approach for $R_{ee}(s)$. When we use, instead, the approaches
that account for the renormalon structure of $R_{ee}(s)$,
and/or we use the value (\ref{alphas2}) instead of (\ref{alphas1})
for $\alpha_s(m^2_{\tau})$, the values of these contributions
change insignificantly. QED corrections are insignificant.}
we obtain
\begin{eqnarray}
10^{11} \times a_{\mu}^{\rm (v.p.)}
&=& 7029 \pm 96 \quad
{\rm for} \ \langle aGG \rangle \ {\rm value \ Eq.~(\ref{aGG1})}
\label{amuthdatfa}
\\
& = & 6995 \pm 91 \quad
{\rm for} \ \langle aGG \rangle \ {\rm value \ Eq.~(\ref{aGG2})}
\ .
\label{amuthdatfb}
\end{eqnarray}
The result obtained in the analysis of Ref.~\cite{Davier:1998si} is
\begin{equation}
10^{11} \times a_{\mu}^{\rm (v.p.)}  = 6924 \pm 62
\quad (\rm D.H. \cite{Davier:1998si}) \ .
\label{amuthdatfDH}
\end{equation}
While the bulk of the result (\ref{amuthdatfDH})
of Ref.~\cite{Davier:1998si} was obtained
by taking into account the data on $e^+ e^-$ and the $\tau$
decays, the bulk of the result (\ref{amuthdatfa})--(\ref{amuthdatfb})
is obtained here by careful resummation of the contour
integral of (\ref{dec2}) where we account for the
renormalon structure of the associated Adler function
and for the $d > 0$ terms.

\section{Comparisons}

How do our results compare with recent results of others
on $a_{\mu}^{\rm (v.p.)}$?

In Table \ref{tabl1} we show the values of
$a_{\mu}^{\rm (v.p.)}$ as predicted 
by others \cite{ADH,BW,J,N1N2,TY1TY2,Davier:1998si}, 
along with our values.\footnote{
We did not include in the Table the results of some earlier
analyses \cite{earlier}.}
We see that our values are consistent with the values obtained by
most of the other authors. We recall the fact that the bulk
of our result is obtained by the pQCD$+$renormalon
calculation, while the bulk of the results of the
others was obtained by data integration.
We observe that the consistency with the results of the others
is even slightly stronger when we take the value of the gluon 
condensate of Eq.~(\ref{aGG2}),
i.e., the small gluon condensate value obtained by the
ALEPH analysis \cite{Barate:1998uf} and suggested also
by the renormalon formalism of \cite{Lee:2001ws}.
Several of the entries in the Table are based on
inclusion of the $\tau$ decay data.\footnote{
The entries of Table \ref{tabl1} are based on the use of
the dispersion relation (\ref{di}). The latter has been questioned 
in Ref.~\cite{Uretsky:2001nx}, because there is no proof that
the photon propagator is at least polynomially bounded.
The inclusion of the $\tau$ decay data for calculation of 
$a_{\mu}^{\rm (v.p.)}$ has been questioned in 
Ref.~\cite{Melnikov:2001uw}, and the difficulties connected with 
this inclusion have been investigated in Ref.~\cite{Cirigliano:2001er}.}

How do our results (\ref{amuthdatfa})--(\ref{amuthdatfb})
compare with the experimental predictions for $a_{\mu}$?
This question remains somewhat unclear due to
theoretical uncertainties of several higher order
hadronic contributions, as argued by the authors
of Ref.~\cite{TY1TY2}. The largest theoretical uncertainty
is in the calculation of the hadronic light--by--light (l.l.)
contributions. The chiral model (ch.m.) approaches would predict
$10^{11} \times a_{\mu}^{\rm (l.l)} = -86 \pm 25$
\cite{Bijnens:1996xf}; the quark constituent model (q.c.m.)
would predict $+92 \pm 20$ % \cite{Laporta:1993pa,TY1TY2}.
\cite{TY1TY2}.
The quark constituent model is valid for large
values of the virtual photon momenta only, so
we will give results for the chiral model unless
otherwise stated.
The QED corrections 
of the type of Fig.~\ref{fig2}
have already been included in (\ref{amuthdatfa})--(\ref{amuthdatfb}).
In addition, there are other QED radiative corrections
where the photon propagator does not have both ends
attached to the hadronic blob (the ``rest'')
$10^{11} \times a_{\mu}({\rm rad.corr., rest})
=-101 \pm 6$ \cite{Krause:1997rf}.
These two radiative corrections $(-86 \pm 26)$ and $(-101 \pm 6)$ 
have to be added to the entries of Table \ref{tabl1}
to obtain the total hadronic contribution
$a_{\mu}^{\rm (hadr.)}$. In our case, we add them
to the  hadronic vacuum polarization 
contribution (\ref{amuthdatfa})--(\ref{amuthdatfb})
(we add the uncertainties in quadrature), and obtain
\begin{eqnarray}
10^{11} \times a_{\mu}^{\rm (hadr.)}
&=& 6842 \pm 100 \quad
{\rm for} \ \langle aGG \rangle \ {\rm value \ Eq.~(\ref{aGG1}) }
\label{amuhchma}
\\
&=& 6808 \pm 95 \quad
{\rm for} \ \langle aGG \rangle \ {\rm value \ Eq.~(\ref{aGG2}) }
\ .
\label{amuhchmb}
\end{eqnarray}
If we took, instead, the quark constituent model result for the 
light--by--light contributions, we would obtain 
$7014 \pm 98$ and $6980 \pm 93$, respectively.

We can now add the results (\ref{amuhchma})--(\ref{amuhchmb})
to the well known electroweak contribution 
(\cite{Hughes:1999fp,Marciano:2001qq} and references therein)
\begin{equation}
10^{11} \times a_{\mu}({\rm EW}) = 116 \ 584 \ 858 \pm 5
\label{EW}
\end{equation}
and obtain the predictions
\begin{eqnarray}
10^{11} \times a_{\mu}({\rm predicted}) &=&
116 \ 591 \ 700 \pm 100 \quad
{\rm for} \ \langle aGG \rangle \ {\rm value \ Eq.~(\ref{aGG1}) }
\label{amuprchma}
\\
&=&  116 \ 591 \ 666 \pm 95 \quad
{\rm for} \ \langle aGG \rangle \ {\rm value \ Eq.~(\ref{aGG2}) }
\ .
\label{amuprchmb}
\end{eqnarray}
The actual experimental number \cite{Brown:2001mg},
when averaged with the older measurements \cite{Bailey:1979mn}, is
\begin{equation}
10^{11} \times a_{\mu}({\rm experiment, \ averaged}) =
116 \ 592 \ 030 \pm 152 \ .
\label{amuexp}
\end{equation}
The predictions (\ref{amuprchma}),  (\ref{amuprchmb})
thus differ from the experimental result by
$330 \pm 182$ ($1.81 \sigma$) and 
$364 \pm 179$ ($2.03 \sigma$),
respectively, where $\sigma \!=\! 182, 179$ is obtained by
adding in quadrature the experimental ($\sigma_{\rm exp.}\!=\!152$)
and the theoretical uncertainty 
($\sigma_{\rm th.}\!=\!100, 95$).\footnote{
The decrease of $C_1$ to zero would contribute, in
quadrature, additional $97$ and $63$ units to the
uncertainties, respectively -- as mentioned in the
previous Section. In such a case, the predictions
would deviate from the experimental result by
$330 \pm 206$ and $364 \pm 190$, respectively.}
If we took, instead of the chiral model result, the
quark constituent model result for the light--by--light
contributions, the corresponding deviations would be
$152 \pm 181$ ($0.84 \sigma$) and 
$186 \pm 178$ ($1.04 \sigma$), respectively.

If we take only the newest experimental number \cite{Brown:2001mg}
\begin{equation}
10^{11} \times a_{\mu}({\rm experiment, \ new}) =
116 \ 592 \ 020 \pm 152 \ ,
\label{amuexpnew}
\end{equation}
then the predictions (\ref{amuprchma})--(\ref{amuprchmb})
differ from it by
$320 \pm 182$ ($1.76 \sigma$) and 
$354 \pm 179$ ($1.98 \sigma$),
respectively. These deviations would be
$0.78 \sigma$ and $0.99 \sigma$, respectively,
if we used the quark constituent model results for the
light--by--light contributions.

\section{Summary}

We obtain clear deviations of the theoretical results
from the experimental ones when we use the chiral approach
results for the light--by--light contributions. 
However, these deviations
nonetheless are significantly smaller than
the $430 \pm 165$ ($ 2.6 \sigma$) difference \cite{Brown:2001mg}\footnote{
There, $\sigma\approx 165$ was obtained by adding in quadrature
$\sigma_{\rm exp.} \approx 150$ and $\sigma_{\rm th.} \approx 67$,
where the value of $\sigma_{\rm th.}$
was taken from Ref.~\cite{Davier:1998si}.}
that has been suggested as the appearance of new physics.
The deviations of our predictions 
from the new experimental values (\ref{amuexp}), (\ref{amuexpnew}),
range from $320 \pm 182$ ($1.76 \sigma$) to
$364 \pm 179$ ($2.03 \sigma$), 
depending mostly on the taken values of the gluon condensate.  
Our results are consistent with those of most
of the other authors who used data integration for the bulk of 
their results. In contrast, the bulk of our results was obtained
by pQCD$+$renormalon calculation.

A major contribution to the aforementioned
reduction of the deviations are
our values for the 
hadronic vacuum polarization 
contributions (CLS1 and CLS2 in Table \ref{tabl1}),
which are significantly higher than those of
Davier and H\"ocker \cite{Davier:1998si}
(DH \cite{Davier:1998si} in Table \ref{tabl1}).
Our values originate from 
resummation of a major part of this contribution via
a contour integral in the complex energy plane
and accounting for the renormalon structure of
the associated Adler function in the integrand.
The accounting for the renormalon structure
contributes about 50 units to $10^{11} \times a_{\mu}^{\rm (v.p.)}$,
and this is larger than the expected future experimental uncertainties.

\acknowledgments

The work of G.C. and I.S. was supported by FONDECYT (Chile),
Grant. No. 1010094 and 8000017, respectively.
The work of T.L. was supported by BK21 Core Project.
We thank M.~Davier, A.~H\"ocker and F.J.~Yndur\'ain
for helpful discussions.

\noindent

\noindent

\begin{table}[ht]
%\vspace{0.2cm}
\par
\begin{center}
\begin{tabular}{l l  c }
 authors & $10^{11} \times a_{\mu}^{\rm ({v.p.})}$ & method \\
\hline \hline     
ADH \cite{ADH} & $7011 \pm 94$ & $e^+e^- + \tau$ data \\
BW \cite{BW}   & $7026 \pm 160$ & $e^+e^-$ data        \\
J \cite{J}     & $6974 \pm 105$ &  mostly $e^+e^-$ data \\
N1 \cite{N1N2} & $7031 \pm 77$  & $e^+e^- + \tau$ data \\
N2 \cite{N1N2} & $7011 \pm 117$ & $e^+e^-$ data \\
TY1 \cite{TY1TY2} & $7002 \pm 65$ &
$e^+e^- + \tau + {\rm spacel.} F_{\pi}(t)$ data \\
TY2 \cite{TY1TY2} & $6982 \pm 97$ &
$e^+e^- + {\rm spacel.} F_{\pi}(t)$ data \\
DH \cite{Davier:1998si} & $6924 \pm 62$ & 
mostly $e^+e^- + \tau$ data \\
CLS1     & $7029 \pm 96$ & mostly theory, and (\ref{aGG1}) value \\
CLS2     & $6995 \pm 91$ & mostly theory, and (\ref{aGG2}) value
\end{tabular}
\end{center}
%\vspace{-0.4cm}
\caption {\footnotesize Comparison of predictions of the hadronic 
vacuum polarization contributions to $a_{\mu}$ by various authors.}
\label{tabl1}
\end{table}

\end{document}